\documentclass[runningheads]{llncs}

\usepackage[T1]{fontenc}
\usepackage[utf8]{inputenc}
\usepackage{graphicx}
\usepackage{booktabs}
\usepackage{amsmath}
\usepackage{multirow}
\usepackage{url}
\usepackage{float}
\usepackage{tikz}
\usepackage{hyperref}
\usetikzlibrary{arrows.meta,positioning,calc}

\begin{document}

\title{Scale-Aware 3D Deep Learning for Robust Brain Metastasis Detection in Multimodal MRI}

\titlerunning{Scale-Aware 3D Deep Learning for Robust Brain Metastasis Detection}

\author{Sylvain Jaume, Hongming Wang, Simon K. Warfield}
\authorrunning{Sylvain Jaume, Hongming Wang, Simon K. Warfield}

\institute{Massachusetts Institute of Technology\\
    Harvard University\\
    Boston Children's Hospital, Harvard Medical School\\\email{sylvain@csail.mit.edu}}
\maketitle

\begin{abstract}
Detecting brain metastases in magnetic resonance imaging (MRI) remains challenging because lesions vary widely in size and appearance, with very small metastases occupying only a minute fraction of a three-dimensional input. We investigate whether combining different spatial fields of view (FOVs) improves lesion detection in multimodal MRI and present a scale-aware 3D deep-learning framework.  The method uses independently trained $96^3$ and $64^3$ 3D U-Nets whose whole-volume probability maps are combined by weighted late fusion. This design allows us to study the effect of spatial context separately from image resolution and modality choice.
On a 97-patient development cohort, cross-FOV fusion improved lesion-level precision and F1 while substantially reducing false positives relative to the individual models. A same-FOV ensemble control showed that these gains were not explained solely by averaging independently trained networks, supporting a contribution from complementary spatial context.
An exploratory cross-FOV agreement filter reduced false positives but did not improve overall F1. These results support cross-FOV probability fusion as a simple and computationally practical strategy for improving the precision–false-positive trade-off in 3D brain-metastasis detection.

\end{abstract}

\section{Introduction}

Brain metastases vary widely in size and location, creating severe spatial-scale imbalance for 3D networks. Large fields of view (FOVs) provide context but reduce the relative representation of small lesions, whereas smaller FOVs increase lesion occupancy while sacrificing context. U-Net and nnU-Net establish strong volumetric baselines and the importance of patch configuration \cite{unet,unet3d,nnunet}. Earlier multiresolution medical-image analysis addressed registration, shape characterization, anatomical labeling, connectome segmentation, and tumor delineation \cite{jaume2001registration,jaume2001shape,jaume2002labeling,jaume2011multiscale,kaus2001}.

Small-lesion detection remains difficult in brain metastasis analysis; prior work has used lesion-specific training, longitudinal information, multi-scale features, cascades, cropped regions, and feature fusion \cite{bousabarah2020,hammer2024,yin2022,amemiya2022,dikici2020,yoo2021,yoo2022,li2023}. Multi-resolution and ensemble strategies are also established in brain-tumor segmentation \cite{soltaninejad2021,ahmad2022,hua2019,kamnitsas2018}. We therefore study a narrower controlled question: how sampling, spatial FOV, and post-hoc probability fusion affect lesion-level detection when the backbone is fixed. Compact 3D U-Nets operate on $96^3$ and $64^3$ inputs, with sensitivity stratified by lesion volume, operating-point analyses, and independent BraTS 2026 validation.

\section{Methods}

\subsection{Dataset}

Experiments used the BraTS 2026 Brain Metastases (BraTS-METS) dataset, comprising multi-institutional pre- and post-treatment multiparametric MRI \cite{maleki2025bratsmets,bratsmet2023}. In accordance with Task~1 rules, models used only the provided data and no pretrained segmentation weights.

Our fixed internal development cohort (DEV97) contained 97 patients and 667 connected-component ground-truth lesions. Patch optimization used a seed-42 90/10 split created separately within positive and negative patch sets. Because this split was patch-level rather than patient-level, patch validation is not treated as evidence of patient-independent generalization; whole-volume lesion results are reported on DEV97.

Inputs were T1c and T2-FLAIR, chosen as complementary tumor/surrounding-signal contrasts while keeping computation tractable; no modality ablation was performed. BraTS-METS preprocessing provides 1-mm isotropic sampling, so $96^3$ and $64^3$ inputs span approximately $96^3$ and $64^3$~mm$^3$, respectively \cite{maleki2025bratsmets,bratsmet2023}. Patch extraction changes spatial extent but performs no additional resampling; both branches use the same 1-mm grid. Ground-truth lesions are connected components of the tumor segmentation, with physical volume computed from voxel count.

\subsection{Lesion-Scale Stratification}

Lesions were stratified by volume into $<0.010$, 0.010--0.025, 0.025--0.050, 0.050--0.100, 0.100--0.500, and $\geq0.500$~mL, containing 70, 95, 87, 84, 139, and 192 lesions, respectively. This stratification was fixed before evaluation.

\subsection{Baseline 3D Network}

\begin{figure}[t]
\centering
\resizebox{0.98\textwidth}{!}{%
\begin{tikzpicture}[
    font=\scriptsize,
    >=Latex,
    node distance=0.55cm and 0.75cm,
    box/.style={draw, rounded corners, align=center, minimum height=0.95cm, minimum width=2.15cm},
    input/.style={box, fill=gray!12},
    model/.style={box, fill=blue!10},
    recon/.style={box, fill=green!10},
    fusion/.style={box, fill=orange!18},
    post/.style={box, fill=purple!10},
    output/.style={box, fill=red!10},
    flow/.style={draw,->,thick},
    branch/.style={draw,->,thick,dashed}
]
\node[input] (mri) {Input MRI\\T1c + T2-FLAIR\\1 mm isotropic};
\node[input, right=of mri] (sw) {Sliding-window\\patch extraction\\and inference};

\node[model, above right=1.0cm and 0.9cm of sw] (f96) {Branch A\\3D U-Net\\$96^3$ patches};
\node[model, below right=1.0cm and 0.9cm of sw] (f64) {Branch B\\3D U-Net\\$64^3$ patches};

\node[recon, right=of f96] (p96) {Reconstruct\\whole-volume\\probability map $P_{96}$};
\node[recon, right=of f64] (p64) {Reconstruct\\whole-volume\\probability map $P_{64}$};

\node[fusion, right=2cm of $(p96)!0.5!(p64)$] (fuse) {Late fusion\\$P = 0.60\,P_{96} + 0.40\,P_{64}$};
\node[post, right=of fuse] (post) {Thresholding +\\connected-component filtering\\$\tau = 0.60$, min volume $=0.010$ mL};
\node[output, right=of post] (out) {Final lesion mask\\BraTS Task 1 export\\foreground $\rightarrow$ label 3};

\draw[flow] (mri) -- (sw);
\draw[branch] (sw) -- (f96);
\draw[branch] (sw) -- (f64);
\draw[flow] (f96) -- (p96);
\draw[flow] (f64) -- (p64);
\draw[flow] (p96.east) -- ++(0.25,0) |- (fuse.west);
\draw[flow] (p64.east) -- ++(0.25,0) |- (fuse.west);
\draw[flow] (fuse) -- (post);
\draw[flow] (post) -- (out);

\node[align=center, anchor=north] at ($(f96.north)!0.5!(p96.north)+(0,0.45)$) {Large-context branch};
\node[align=center, anchor=south] at ($(f64.south)!0.5!(p64.south)+(0,-0.45)$) {Small-context branch};
\end{tikzpicture}%
}
\caption{Overview of the proposed dual-field-of-view detection pipeline. T1c and T2-FLAIR MR images are processed with sliding-window inference using two independently trained 3D U-Nets with input sizes $96^3$ and $64^3$. Their reconstructed whole-volume probability maps are combined by weighted late fusion, followed by thresholding and connected-component filtering to obtain the final lesion prediction used for BraTS Task~1 submission.}
\label{fig:method_pipeline}
\end{figure}
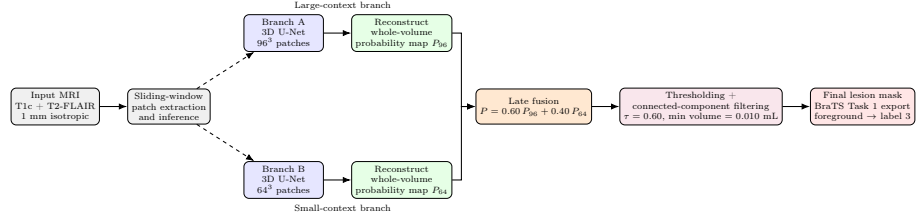

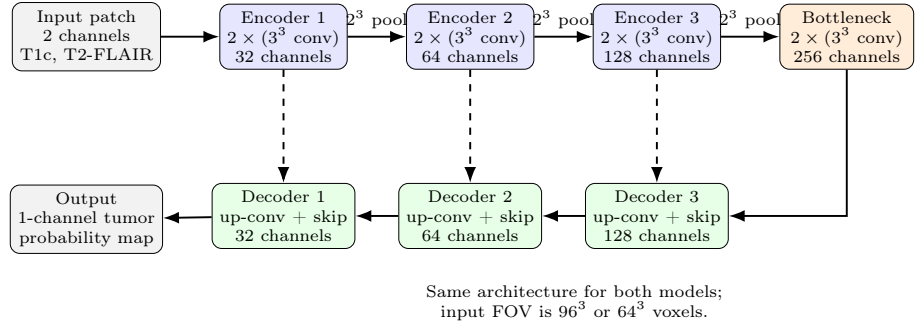
\begin{figure}[t]
\centering
\resizebox{0.98\textwidth}{!}{%
\begin{tikzpicture}[
    font=\scriptsize,
    >=Latex,
    node distance=0.65cm and 0.85cm,
    enc/.style={draw, rounded corners, fill=blue!10, minimum width=1.85cm, minimum height=0.82cm, align=center},
    dec/.style={draw, rounded corners, fill=green!10, minimum width=1.85cm, minimum height=0.82cm, align=center},
    bott/.style={draw, rounded corners, fill=orange!15, minimum width=1.95cm, minimum height=0.88cm, align=center},
    io/.style={draw, rounded corners, fill=gray!10, minimum width=1.85cm, minimum height=0.82cm, align=center},
    skip/.style={draw,->,thick,dashed},
    flow/.style={draw,->,thick}
]
\node[io] (input) {Input patch\\2 channels\\T1c, T2-FLAIR};
\node[enc, right=of input] (enc1) {Encoder 1\\$2\times(3^3$ conv$)$\\32 channels};
\node[enc, right=of enc1] (enc2) {Encoder 2\\$2\times(3^3$ conv$)$\\64 channels};
\node[enc, right=of enc2] (enc3) {Encoder 3\\$2\times(3^3$ conv$)$\\128 channels};
\node[bott, right=of enc3] (bott) {Bottleneck\\$2\times(3^3$ conv$)$\\256 channels};
\node[dec, below=1.65cm of enc3] (dec3) {Decoder 3\\up-conv + skip\\128 channels};
\node[dec, below=1.65cm of enc2] (dec2) {Decoder 2\\up-conv + skip\\64 channels};
\node[dec, below=1.65cm of enc1] (dec1) {Decoder 1\\up-conv + skip\\32 channels};
\node[io, below=1.65cm of input] (out) {Output\\1-channel tumor\\probability map};
\draw[flow] (input) -- (enc1);
\draw[flow] (enc1) -- node[above] {$2^3$ pool} (enc2);
\draw[flow] (enc2) -- node[above] {$2^3$ pool} (enc3);
\draw[flow] (enc3) -- node[above] {$2^3$ pool} (bott);
\draw[flow] (bott) |- (dec3);
\draw[flow] (dec3) -- (dec2);
\draw[flow] (dec2) -- (dec1);
\draw[flow] (dec1) -- (out);
\draw[skip] (enc3.south) -- (dec3.north);
\draw[skip] (enc2.south) -- (dec2.north);
\draw[skip] (enc1.south) -- (dec1.north);
\node[align=center, below=0.42cm of dec3, xshift=-1.2cm] {Same architecture for both models;\\input FOV is $96^3$ or $64^3$ voxels.};
\end{tikzpicture}%
}
\caption{Compact 3D U-Net architecture used for both field-of-view models. The network takes two MRI channels (T1c and T2-FLAIR), uses three encoder levels with 32, 64, and 128 channels and a 256-channel bottleneck, and reconstructs a one-channel tumor-probability map through a symmetric decoder with skip connections. The $96^3$ and $64^3$ models use the same architecture and differ in input field of view.}
\label{fig:unet_architecture}
\end{figure}

Figures~\ref{fig:method_pipeline} and \ref{fig:unet_architecture} summarize the dual-FOV workflow and shared compact 3D U-Net. The encoder uses 32/64/128 channels and a 256-channel bottleneck, with two $3^3$ convolutions per level, instance normalization, ReLU, and $2^3$ max pooling; the decoder uses transposed convolutions and skip connections. A final $1^3$ convolution produces a binary tumor logit map. Decoder upsampling restores predictions to the 1-mm input grid. The two branches use the same architecture and training procedure and differ in their \(96^3\) versus \(64^3\) input FOV.

\subsection{Patch-Based Training}

Positive patches are centered on annotated lesions and negatives on non-lesional brain regions, with approximately balanced sampling. Both FOV models use Dice--focal loss and the same optimization procedure. Adam was used at $10^{-4}$ with batch size 2 for up to 50 epochs; ReduceLROnPlateau reduced the learning rate by 0.5 after 5 stagnant validation epochs. Training used mixed precision when CUDA was available and independent axis flips ($p=0.5$); the best validation-Dice checkpoint was retained. Experiments ran on an AMD Ryzen 7 5800H, 16~GB RAM, and RTX 3060 GPU (6~GB). The five-epoch seed-43 $64^3$ control required 51.9 minutes.
The baseline field of view is $96^3$ voxels.

The independently initialized seed-43 \(64^3\) control was trained for five epochs rather than the 50-epoch maximum used for the primary models. Its best observed validation Dice was reached at epoch 5 and the corresponding checkpoint was retained. This control was introduced to test whether gains from the \(96^3+64^3\) ensemble could be explained solely by averaging independently trained models, rather than to provide a fully matched retraining of the primary \(64^3\) branch. The shorter training schedule may therefore underestimate the performance achievable by a fully optimized same-FOV ensemble.

\subsection{Small-Lesion-Aware Sampling}

To isolate sampling frequency from FOV, positive patches were weighted 4, 3, 2, and 1 for lesion volumes $<0.025$, 0.025--0.050, 0.050--0.100, and $\geq0.100$~mL, respectively; negatives received unit weight, with normalization preserving approximately balanced positive/negative sampling. In 5,000 sampled patches, 2,504 were positive and 2,496 negative, and 55.6\% of positives represented lesions $<0.025$~mL.

\subsection{Small-Field-of-View Model}

The $64^3$ model uses the same architecture and optimization as the baseline, with inputs obtained by center-cropping the lesion-centered $96^3$ patches. Thus, the lesion occupies a larger fraction of the input while the underlying training lesion is unchanged, isolating FOV as the principal experimental variable.

\subsection{Scale-Aware Multi-Field-of-View Inference}

For the internal development experiments, our method combines probability maps generated from separately trained $96^3$ and $64^3$ networks. Unlike feature-level multi-scale architectures, the two networks are optimized separately and interact only after whole-volume probability reconstruction. This design is intentional: it permits direct ablation of field of view and post-hoc fusion without changing either backbone.

Each network is applied independently to the complete MRI volume using overlapping sliding-window inference with 50\% overlap in each spatial dimension. Thus, the $96^3$ model uses a stride of 48 voxels and the $64^3$ model a stride of 32 voxels. Overlapping patch predictions are aggregated to produce spatially aligned whole-volume probability maps $P_{96}(\mathbf{x})$ and $P_{64}(\mathbf{x})$, where $\mathbf{x}$ denotes a voxel location.

The two probability maps can be fused using a weighted combination,
\[
P_{\mathrm{fused}}(\mathbf{x})
=
\alpha P_{96}(\mathbf{x})
+
(1-\alpha)P_{64}(\mathbf{x}),
\]
where $\alpha \in [0,1]$ is the weight assigned to the $96^3$ large-context predictor and $1-\alpha$ is the weight assigned to the $64^3$ predictor.

Because both maps are reconstructed independently, fusion parameters can be evaluated without retraining. No explicit probability-calibration procedure was applied before fusion; consequently, the optimized value of $\alpha$ may reflect both complementary prediction errors and differences in probability calibration between the two separately trained internal models.

\subsection{Whole-Volume Lesion Extraction}

The resulting probability map is thresholded to obtain a binary tumor prediction. Connected-component analysis is then used to identify individual predicted lesions.

Components below a configurable minimum physical volume are removed. For each remaining lesion, the system records its volume, centroid, bounding box, and voxel count.

Probability threshold and minimum-component-volume parameters are swept over the internal development cohort rather than selecting a single operating point a priori. The selected operating points should therefore be interpreted as development-set choices rather than unbiased test estimates. The 0.010~mL component filter used in the primary fusion comparison may disadvantage predictions corresponding to the smallest reference-lesion stratum ($<0.010$~mL); this trade-off is addressed explicitly in the Discussion.

\subsection{Lesion Matching and Evaluation}

Predicted and reference components are matched greedily one-to-one at IoU $\geq0.10$: for each reference lesion, the unmatched prediction with highest IoU is accepted if it meets the threshold, and cannot be reused. Matched pairs are TP, unmatched predictions FP, and unmatched references FN. We report sensitivity $=\mathrm{TP}/(\mathrm{TP}+\mathrm{FN})$, precision $=\mathrm{TP}/(\mathrm{TP}+\mathrm{FP})$, their harmonic mean (F1), and FP/patient. For the multi-FOV ablation, all models use the same 0.010-mL minimum component volume while the probability threshold is swept. Internal lesion-level metrics are development measures and are not numerically interchangeable with the official BraTS instance-F1 computed on a separate cohort and evaluation pipeline.

\subsection{Reproducibility}
Code is available at \url{https://github.com/ailabrepo/BraTSDetector} . The retained $64^3$ checkpoint had patch training/validation Dice 0.686/0.785; the seed-43 control reached 0.692/0.775 in 51.9 minutes. The original \(96^3\) checkpoint file was not retained, although its whole-volume probability maps had been generated and stored before the later checkpoint overwrite. The challenge container contained identical learned weights under both checkpoint filenames and thus applied one retained model at two inference FOVs; the official score is therefore not an external replication of the separately trained internal ensemble.

\section{Results}

\subsection{Sampling Ablation}

Small-lesion-aware sampling increased sensitivity at some fixed thresholds but also increased false positives. At approximately matched burden, baseline sensitivity was 0.544 at 4.00 FP/patient versus 0.501 at 3.99 FP/patient with weighted sampling, with no consistent size-stratified advantage. Thus, increased sampling frequency alone was insufficient.

\subsection{Effect of Reducing the Field of View}

The $64^3$ model tests whether a smaller FOV is beneficial in isolation. Its best observed F1 was 0.413 at threshold 0.70 (sensitivity 0.484, precision 0.360, 5.907 FP/patient). An independently trained $64^3$ replicate (seed 43) reached F1 0.433 at the same threshold (sensitivity 0.541, precision 0.361, 6.588 FP/patient). Thus, reducing FOV alone did not outperform the $96^3$ baseline.

\subsection{Multi-Field-of-View Fusion}

Weighted probability fusion produced substantially better lesion-level operating points than either single-FOV endpoint. Table~\ref{tab:fusionablation} summarizes the best-F1 operating point observed across the fusion-weight sweep, using the same minimum component volume of 0.010~mL. Here, $\alpha$ denotes the weight of the $96^3$ probability map.

\begin{table}[t]
\centering
\caption{Fusion-weight ablation at a common minimum component volume of 0.010~mL. Each row reports the probability threshold giving the highest observed lesion-level F1 for that fusion weight. The $\alpha=0$ and $\alpha=1$ rows correspond to the $64^3$ and $96^3$ single-FOV endpoints, respectively.}
\label{tab:fusionablation}
\begin{tabular}{rrrrrr}
\toprule
$\alpha$ & Threshold & Sensitivity & Precision & F1 & FP/patient \\
\midrule
0.00 & 0.70 & 0.484 & 0.360 & 0.413 & 5.907 \\
0.25 & 0.80 & 0.421 & 0.603 & 0.496 & 1.907 \\
0.50 & 0.50 & 0.615 & 0.560 & 0.586 & 3.320 \\
0.60 & 0.60 & 0.582 & 0.602 & \textbf{0.592} & 2.639 \\
0.70 & 0.70 & 0.543 & \textbf{0.651} & \textbf{0.592} & 2.000 \\
0.80 & 0.80 & 0.495 & 0.627 & 0.553 & 2.021 \\
1.00 & 0.50 & \textbf{0.640} & 0.417 & 0.505 & 6.165 \\
\bottomrule
\end{tabular}
\end{table}

Intermediate fusion weights improved the precision--sensitivity balance. At $\alpha=0.60$, fusion achieved sensitivity 0.582, precision 0.602, F1 0.592, and 2.639 FP/patient, a 17.2\% relative F1 increase and 57.2\% FP/patient reduction versus the $96^3$ endpoint. Patient-level bootstrap analysis (10,000 resamples) gave F1 0.592 (95\% CI 0.538--0.643) versus 0.505 (95\% CI 0.428--0.583) for the $96^3$ model; the paired F1 difference was +0.087 (95\% CI 0.024--0.150; $p=0.003$), with 3.526 fewer FP/patient (95\% CI 1.691--5.722). We selected $\alpha=0.60$ over the equally rounded-F1 $\alpha=0.70$ setting because it detected 26 additional true lesions.

\subsection{Same-FOV Ensemble Control}

To distinguish generic ensemble averaging from an FOV-diversity effect, we trained an independent second $64^3$ model (seed 43) and averaged its probability map 50/50 with the original $64^3$ model. At each method's best-F1 threshold, the same-FOV $64^3+64^3$ ensemble achieved sensitivity 0.523, precision 0.472, F1 0.496, and 4.021 FP/patient. The cross-FOV $96^3+64^3$ ensemble achieved sensitivity 0.559, precision 0.630, F1 0.593, and 2.258 FP/patient. In a paired patient-level bootstrap, cross-FOV fusion improved F1 by 0.096 (95\% CI 0.059--0.132; $p<0.001$), improved precision by 0.158 (95\% CI 0.113--0.203; $p<0.001$), and reduced FP/patient by 1.763 (95\% CI 1.247--2.340 fewer FP/patient; $p<0.001$). The sensitivity difference (+0.036) was not significant (95\% CI $-0.008$--0.075; $p=0.114$). Thus, generic ensembling contributes to performance, while the observed cross-FOV results provide evidence of an additional precision/F1 and false-positive benefit.

\subsection{Lesion-Size Performance of the Selected Fusion Model}

We next evaluated lesion sensitivity as a function of ground-truth lesion volume using the selected $\alpha=0.60$ fusion, probability threshold 0.60, and minimum predicted-component volume of 0.010~mL. Table~\ref{tab:sizefinal} reports the results.

\begin{table}[t]
\centering
\caption{Lesion-size-stratified sensitivity for the selected $\alpha=0.60$ fusion model at probability threshold 0.60 and minimum predicted-component volume 0.010~mL.}
\label{tab:sizefinal}
\begin{tabular}{lrrr}
\toprule
Lesion volume (mL) & N & Detected & Sensitivity \\
\midrule
$<0.010$       & 70  & 4   & 0.057 \\
$0.010$--$0.025$ & 95  & 31  & 0.326 \\
$0.025$--$0.050$ & 87  & 36  & 0.414 \\
$0.050$--$0.100$ & 84  & 49  & 0.583 \\
$0.100$--$0.500$ & 139 & 97  & 0.698 \\
$\geq0.500$      & 192 & 171 & 0.891 \\
\bottomrule
\end{tabular}
\end{table}

Sensitivity increased monotonically from 0.057 below 0.010~mL to 0.891 at $\geq0.500$~mL. Under matched post-processing, fusion did not improve sensitivity within individual size strata relative to the $96^3$ baseline; its main benefit was improved precision/F1 through false-positive suppression.

\subsection{Operating-Point Robustness}

We evaluated the effect of the minimum predicted-component volume while holding the selected fusion weight ($\alpha=0.60$) and probability threshold (0.60) fixed. Among thresholds of 0, 0.0025, 0.005, 0.010, 0.025, and 0.050~mL, the selected 0.010-mL filter yielded the highest overall F1 (0.592). Removing component filtering increased overall sensitivity from 0.582 to 0.619 and sensitivity for lesions $<0.010$~mL from 0.057 to 0.171, but increased FP/patient from 2.64 to 6.32 and reduced F1 to 0.488. Increasing the filter to 0.025 or 0.050~mL further reduced FP/patient (1.69 and 1.10) but also reduced F1 (0.572 and 0.533). Thus, the 0.010-mL setting represents an empirical trade-off between very-small-lesion sensitivity and false-positive suppression.

\begin{figure}[t]
\centering
\includegraphics[width=0.72\textwidth]{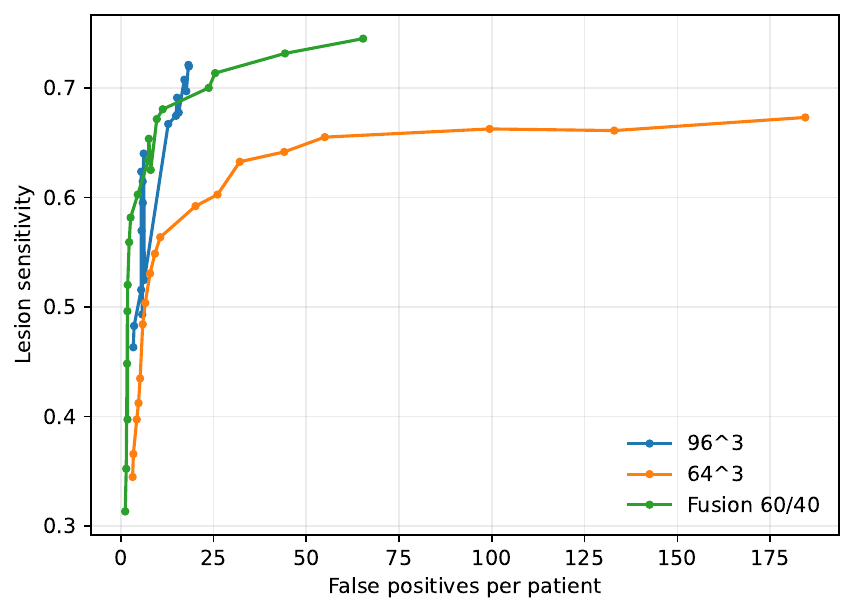}
\caption{Lesion-level FROC analysis on the 97-patient internal development cohort for the $96^3$, $64^3$, and 60/40 fusion models, using a fixed minimum predicted-component volume of 0.010~mL. Probability thresholds were swept from 0.10 to 0.95. The fused model provides a more favorable sensitivity--false-positive trade-off in the low-false-positive operating range and achieves its highest F1 of 0.593 at threshold 0.65.}
\label{fig:froc}
\end{figure}

We also performed a lesion-level FROC sweep over probability thresholds for the $96^3$, $64^3$, and 60/40 fusion models (Fig.~\ref{fig:froc}). When each method was allowed to select its own best-F1 probability threshold, fusion reached F1 0.593 at threshold 0.65 with sensitivity 0.559 and 2.26 FP/patient. The $96^3$ model reached F1 0.515 at threshold 0.55 with sensitivity 0.624 and 5.49 FP/patient, while the $64^3$ model reached F1 0.413 at threshold 0.70. The fusion advantage therefore persists across the probability-threshold sweep rather than depending on a single baseline threshold.

As an exploratory post-hoc false-positive control, we retained the selected 60/40 fusion at probability threshold 0.60 and required each connected-component candidate to receive minimum mean support from both FOV branches. At an agreement threshold of 0.35, FP/patient decreased from 2.639 to 2.268 and precision increased from 0.602 to 0.630, while sensitivity decreased from 0.582 to 0.561. F1 was essentially unchanged (0.592 versus 0.593; paired-bootstrap difference +0.001, 95\% CI $-0.009$--0.012; $p=0.786$). Thus, simple cross-FOV agreement filtering can reduce false positives, but primarily by trading sensitivity rather than improving overall F1.

\subsection{False-Positive Characterization and Challenge Validation}
Using the selected 60/40 fusion at threshold 0.60 with exact greedy one-to-one matching, 256 unmatched predicted components occurred in 65/97 DEV97 patients. We performed an exploratory appearance-based audit of the 40 highest-confidence unmatched components (Table~\ref{tab:fp_categories}). Most were compact focal candidates with lesion-like appearance, while a substantial minority were peripheral or boundary-associated, or large irregular non-lesional predictions. Two cases were near ground truth but remained unmatched because of split/merge or one-to-one matching behavior. This audit is descriptive rather than radiologist-adjudicated, and the proportions refer only to the selected high-confidence subset.

\begin{table}[H]
\centering
\small
\caption{Exploratory appearance categories among the 40 highest-confidence unmatched predictions. Categories are descriptive image-pattern groupings rather than clinical diagnoses.}
\label{tab:fp_categories}
\begin{tabular}{lrr}
\toprule
Appearance category & N & \% \\
\midrule
Compact focal / lesion-like candidate & 23 & 57.5 \\
Peripheral/boundary-associated or anatomical structure & 8 & 20.0 \\
Large diffuse/irregular non-lesional region & 7 & 17.5 \\
Near-GT split/merge or matching case & 2 & 5.0 \\
\bottomrule
\end{tabular}
\end{table}

As an additional diagnostic of false-positive behavior beyond DEV97, we evaluated the independently trained seed-43 \(64^3\) model on all BraTS training-dataset cases for which complete whole-volume predictions were available.. Of the 606 cases with complete predictions, 97 constituted the fixed DEV97 development cohort and the remaining 509 cases formed the non-DEV97 subset used for this analysis. These 509 cases came from the same BraTS-METS training dataset as DEV97 and were not used as a separate held-out test cohort. Because the original pipeline used patch-level rather than patient-level splitting, this non-DEV97 analysis is descriptive and should not be interpreted as an independent held-out evaluation.

The seed-43 \(64^3\) model produced 6.59 false positives per patient on DEV97 and 6.12 false positives per patient on the 509-case non-DEV97 subset, indicating that the substantial false-positive burden was not confined to the development cohort. Thus, substantial FP propensity was also present across the broader training-data population.

On the 179-case BraTS 2026 validation cohort, official ET all-lesion instance F1 was 0.385, lesion-wise DSC 0.155, NSD 0.184, large-lesion F1 0.302, and small-lesion F1 0.105; TC and WT all-lesion F1 were 0.387 and 0.391. Because later checkpoint-file overwriting left identical retained weights under both container checkpoint names, this submission is not an external replication of the separately trained internal ensemble. The binary detector was mapped to ET and does not explicitly predict RC or other subregions.

\section{Discussion}
The same-FOV control used a shorter five-epoch training schedule than the primary models. Although this model reached its best observed validation Dice at epoch 5 and improved the individual-model F1 when ensembled with the original \(64^3\) network, a fully matched 50-epoch control could potentially yield stronger same-FOV performance. The observed advantage of cross-FOV fusion should therefore be interpreted as supportive evidence for complementary spatial context rather than a definitive isolation of field-of-view effects.

Lesion size strongly determined performance: neither oversampling nor the \(64^3\) FOV alone improved the \(96^3\) baseline, and only 4/70 lesions below 0.010~mL were detected by the selected fusion. Cross-FOV fusion instead improved precision and FP burden. The same-FOV control provides evidence of a generic ensemble benefit but does not fully explain the cross-FOV advantage: at best-F1 operating points, \(64^3+64^3\) reached F1 0.496 and 4.021 FP/patient versus 0.593 and 2.258 for cross-FOV fusion; F1 and FP differences were significant, but sensitivity was not.

The FP audit shows both lesion-like mimics and anatomical/boundary errors, and seed-43 FP burden was similar on DEV97 and the broader training-data subset. A post-hoc agreement filter reduced FP/patient from 2.64 to 2.27 without improving F1, motivating hard-negative mining, candidate-level classification, and anatomical constraints. Official validation remained substantially weaker (ET instance F1 0.385), and the submitted challenge container applied the same retained weights at both FOVs.

\section{Conclusion}

We presented a scale-aware 3D deep-learning framework for brain-metastasis detection in multimodal MRI that combines predictions from two spatial fields of view while keeping image resolution and input modalities fixed. The study was designed to isolate the effect of spatial context and probability-level fusion on lesion detection, with particular attention to the precision–sensitivity trade-off in patients with multiple small metastases. On the 97-patient development cohort, weighted fusion of the $96^3$ and $64^3$ probability maps improved precision and lesion-level F1 while substantially reducing false positives relative to either single-FOV model. A same-FOV ($64^3$+$64^3$) control provided evidence that the improvement was not attributable solely to averaging independently trained models: cross-FOV fusion improved F1 by 0.096 and reduced false positives by 1.76 per patient relative to the same-FOV ensemble in paired bootstrap analysis, while the corresponding sensitivity difference was not significant.

Performance increased strongly with lesion size, while detection of very small metastases remained difficult. False-positive auditing showed that residual errors included both compact lesion-like candidates and anatomically implausible or boundary-associated predictions, suggesting that false-positive suppression requires more than simple threshold adjustment. An exploratory cross-FOV agreement filter reduced false positives and increased precision, but at the cost of sensitivity and without a significant improvement in F1. These findings indicate that complementary spatial context is useful for improving lesion-level precision and controlling false positives, but does not by itself solve the small-lesion detection problem.

The official BraTS evaluation also demonstrated a substantial gap between internal and external performance, with an ET instance F1 of 0.385, underscoring the difficulty of robust generalization in this setting. Overall, the results support cross-FOV probability fusion as a simple and computationally practical mechanism for combining complementary spatial context in 3D metastasis detection.

\subsubsection{Disclosure of Interests.}
The authors have no competing interests to declare that are relevant to the content of this article.


\begin{thebibliography}{99}

\bibitem{unet}
Ronneberger, O., Fischer, P., Brox, T.:
U-Net: Convolutional networks for biomedical image segmentation.
In: Medical Image Computing and Computer-Assisted Intervention (MICCAI), pp. 234--241 (2015)

\bibitem{unet3d}
Cicek, O., Abdulkadir, A., Lienkamp, S.S., Brox, T., Ronneberger, O.:
3D U-Net: Learning dense volumetric segmentation from sparse annotation.
In: Medical Image Computing and Computer-Assisted Intervention (MICCAI), pp. 424--432 (2016)

\bibitem{nnunet}
Isensee, F., Jaeger, P.F., Kohl, S.A.A., Petersen, J., Maier-Hein, K.H.:
nnU-Net: A self-configuring method for deep learning-based biomedical image segmentation.
Nature Methods \textbf{18}, 203--211 (2021)

\bibitem{jaume2001registration}
Jaume, S., Ferrant, M., Warfield, S.K., Macq, B.:
Multiresolution parameterization of meshes for improved surface-based registration.
In: Medical Imaging 2001: Image Processing, vol. 4322, pp. 633--642. SPIE (2001)

\bibitem{jaume2001shape}
Jaume, S., Ferrant, M., Schreyer, A., Hoyte, L., Macq, B., Fielding, J., Kikinis, R., Warfield, S.K.:
Multiresolution signal processing on meshes for automatic pathological shape characterization.
In: International Conference on Medical Image Computing and Computer-Assisted Intervention, pp. 1398--1400. Springer (2001)

\bibitem{jaume2002labeling}
Jaume, S., Macq, B., Warfield, S.K.:
Labeling the brain surface using a deformable multiresolution mesh.
In: International Conference on Medical Image Computing and Computer-Assisted Intervention, pp. 451--458. Springer (2002)

\bibitem{jaume2011multiscale}
Jaume, S., Knobe, K., Newton, R.R., Schlimbach, F., Blower, M., Reid, R.C.:
A multiscale parallel computing architecture for automated segmentation of the brain connectome.
IEEE Transactions on Biomedical Engineering \textbf{59}(1), 35--38 (2011)

\bibitem{kaus2001}
Kaus, M.R., Warfield, S.K., Nabavi, A., Black, P.M., Jolesz, F.A., Kikinis, R.:
Automated segmentation of MR images of brain tumors.
Radiology \textbf{218}(2), 586--591 (2001).
doi:10.1148/radiology.218.2.r01fe44586

\bibitem{bousabarah2020}
Bousabarah, K., Ruge, M., Brand, J.-S., et al.:
Deep convolutional neural networks for automated segmentation of brain metastases trained on clinical data.
Radiation Oncology \textbf{15}, 87 (2020).
doi:10.1186/s13014-020-01514-6

\bibitem{hammer2024}
Hammer, Y., Najjar, W., Kahanov, L., Joskowicz, L., et al.:
Two is better than one: longitudinal detection and volumetric evaluation of brain metastases after stereotactic radiosurgery with a deep learning pipeline.
Journal of Neuro-Oncology \textbf{166}, 547--555 (2024)

\bibitem{yin2022}
Yin, S., Luo, X., Yang, Y., et al.:
Development and validation of a deep-learning model for detecting brain metastases on 3D post-contrast MRI: a multi-center multi-reader evaluation study.
Neuro-Oncology \textbf{24}, 1559--1570 (2022).
doi:10.1093/neuonc/noac025

\bibitem{amemiya2022}
Amemiya, S., Takao, H., Kato, S., Yamashita, H., Sakamoto, N., Abe, O.:
Feature-fusion improves MRI single-shot deep learning detection of small brain metastases.
Journal of Neuroimaging \textbf{32}, 111--119 (2022).
doi:10.1111/jon.12916

\bibitem{dikici2020}
Dikici, E., Ryu, J.L., Demirer, M., et al.:
Automated brain metastases detection framework for T1-weighted contrast-enhanced 3D MRI.
IEEE Journal of Biomedical and Health Informatics \textbf{24}, 2883--2893 (2020).
doi:10.1109/JBHI.2020.2982103

\bibitem{yoo2021}
Yoo, Y., Ceccaldi, P., Liu, S., Re, T.J., Cao, Y., Balter, J.M., Gibson, E.:
Evaluating deep learning methods in detecting and segmenting different sizes of brain metastases on 3D post-contrast T1-weighted images.
Journal of Medical Imaging \textbf{8}, 037001 (2021).
doi:10.1117/1.JMI.8.3.037001

\bibitem{yoo2022}
Yoo, S.K., Kim, T.H., Chun, J., et al.:
Deep-learning-based automatic detection and segmentation of brain metastases with small volume for stereotactic ablative radiotherapy.
Cancers \textbf{14}, 2555 (2022).
doi:10.3390/cancers14102555

\bibitem{li2023}
Li, R., Guo, Y., Zhao, Z., et al.:
MRI-based two-stage deep learning model for automatic detection and segmentation of brain metastases.
European Radiology \textbf{33}, 3521--3531 (2023).
doi:10.1007/s00330-023-09420-7

\bibitem{soltaninejad2021}
Soltaninejad, M., Pridmore, T., Pound, M.:
Efficient MRI brain tumor segmentation using multi-resolution encoder-decoder networks.
In: International MICCAI Brainlesion Workshop, pp. 30--39. Springer (2021).
doi:10.1007/978-3-030-72087-2\_3

\bibitem{ahmad2022}
Ahmad, P., Qamar, S., Shen, L., Rizvi, S.Q.A., Ali, A., Chetty, G.:
MS UNet: Multi-scale 3D UNet for brain tumor segmentation.
In: International MICCAI Brainlesion Workshop, pp. 30--41. Springer (2022).
doi:10.1007/978-3-031-09002-8\_3

\bibitem{hua2019}
Hua, R., Huo, Q., Gao, Y., Sun, Y., Shi, F.:
Multimodal brain tumor segmentation using cascaded V-Nets.
In: International MICCAI Brainlesion Workshop, pp. 49--60. Springer (2019).
doi:10.1007/978-3-030-11726-9\_5

\bibitem{kamnitsas2018}
Kamnitsas, K., Bai, W., Ferrante, E., et al.:
Ensembles of multiple models and architectures for robust brain tumour segmentation.
In: International MICCAI Brainlesion Workshop, pp. 450--462. Springer (2018).
doi:10.1007/978-3-319-75238-9\_38

\bibitem{maleki2025bratsmets}
Maleki, N., Amiruddin, R., Moawad, A.W., et al.:
Analysis of the MICCAI Brain Tumor Segmentation--Metastases (BraTS-METS) 2025 Lighthouse Challenge: Brain metastasis segmentation on pre- and post-treatment MRI.
arXiv:2504.12527v3 (2025).
doi:10.48550/arXiv.2504.12527

\bibitem{bratsmet2023}
Moawad, A.W., et al.:
The Brain Tumor Segmentation-Metastases (BraTS-METS) Challenge 2023: Brain metastasis segmentation on pre-treatment MRI.
arXiv:2306.00838 (2023).
doi:10.48550/arXiv.2306.00838

\end{thebibliography}
\end{document}